\documentclass[10pt,conference]{IEEEtran}
\usepackage{tcolorbox}

\usepackage{cite}
\usepackage{parcolumns}
\usepackage{amsmath,amssymb,amsfonts}
\usepackage{algorithmic}
\usepackage{graphicx}
\usepackage{textcomp}
\usepackage{xcolor}
\usepackage{lipsum}
\usepackage{framed}
\usepackage{soul}
\usepackage[export]{adjustbox}
\usepackage{floatrow}
\usepackage{hyperref}
\usepackage{algorithm}
\usepackage{subfiles}
\usepackage{verbatim}
\usepackage{listing}
\usepackage{listings}   
\usepackage{subcaption}
\usepackage{enumitem}
\usepackage{tikz}

\usepackage[T1]{fontenc}
\usepackage{listings,xcolor,beramono}
\usepackage{tikz}
\usepackage{booktabs}
\makeatletter
\newenvironment{btHighlight}[1][]
{\begingroup\tikzset{bt@Highlight@par/.style={#1}}\begin{lrbox}{\@tempboxa}}
{\end{lrbox}\bt@HL@box[bt@Highlight@par]{\@tempboxa}\endgroup}

\newcommand\btHL[1][]{%
  \begin{btHighlight}[#1]\bgroup\aftergroup\bt@HL@endenv%
}
\def\bt@HL@endenv{%
  \end{btHighlight}%
  \egroup
}
\newcommand{\bt@HL@box}[2][]{%
  \tikz[#1]{%
    \pgfpathrectangle{\pgfpoint{1pt}{0pt}}{\pgfpoint{\wd #2}{\ht #2}}%
    \pgfusepath{use as bounding box}%
    \node[anchor=base west, fill=orange!30,outer sep=0pt,inner xsep=1pt, inner ysep=0pt, rounded corners=3pt, minimum height=\ht\strutbox+1pt,#1]{\raisebox{1pt}{\strut}\strut\usebox{#2}};
  }%
}

\newcommand{\code}[1]{\textsf{#1}}

\newcommand{\figref}[1]{Figure~\ref{#1}}

\author{
\IEEEauthorblockN{Ali Shokri}
\IEEEauthorblockA{\textit{Dept. of Software Engineering} \\
\textit{Rochester Institute of Technology}\\ Rochester, NY, USA \\
as8308@rit.edu}
\and
\IEEEauthorblockN{Mehdi Mirakhorli}
\IEEEauthorblockA{\textit{Dept. of Software Engineering} \\
\textit{Rochester Institute of Technology}\\ Rochester, NY, USA \\
mxmvse@rit.edu}

}

\usepackage{tikz}
\usepackage{textcomp}
\usepackage{hyperref}
\usepackage{lipsum}

\makeatother

\lstdefinelanguage{JavaExt}[]{Java}{
  frame=tlrb,
  basicstyle={\tiny\ttfamily},
  morekeywords={assert,constexpr,size_t},
  moredelim=**[is][{\btHL[fill=green!30,draw=red,dashed,thin]}]{@@}{@@},
  moredelim=**[is][{\btHL[fill=lightyellow!90,draw=white,dashed,thin]}]{|}{|},
  moredelim=**[is][{\btHL[fill=azure!90,draw=white,dashed,thin]}]{@|}{|@},
}

\def\BibTeX{{\rm B\kern-.05em{\sc i\kern-.025em b}\kern-.08em
    T\kern-.1667em\lower.7ex\hbox{E}\kern-.125emX}}
    
\definecolor{dkgreen}{rgb}{0,0.6,0}
\definecolor{gray}{rgb}{0.5,0.5,0.5}
\definecolor{mauve}{rgb}{0.58,0,0.82}
\definecolor{lightyellow}{rgb}{1,1,.6}
\definecolor{cyan}{rgb}{0.0, 1.0, 1.0}
\definecolor{azure}{rgb}{0.67, 0.9, 0.93}
\definecolor{carolinablue}{rgb}{0.6, 0.73, 0.89}
\hypersetup{
    colorlinks=true,
    linkcolor=blue,
    filecolor=blue,      
    urlcolor=blue,
}

\newcounter{mycomment}
\newcommand{\mycomment}[2][]{%
\refstepcounter{mycomment}%
{%
\setstretch{0.75}
\todo[inline,color={red!100!green!33},size=\scriptsize]{%
\textbf{[\uppercase{#1}\#\themycomment]:}~#2}%
}}
\newcommand{\task}[1]{\mycomment[Task]{#1}}

\renewcommand{\task}[1]{}

\usepackage{tikz}
\usepackage{textcomp}
\usepackage{hyperref}
\usepackage{lipsum}

\newcommand\copyrighttext{%
  \footnotesize \textcopyright 2021 IEEE. Personal use of this material is permitted. Permission from IEEE must be obtained for all other uses, in any current or future media, including reprinting/republishing this material for advertising or promotional purposes, creating new collective works, for resale or redistribution to servers or lists, or reuse of any copyrighted component of this work in other works.}
\newcommand\copyrightnotice{%
\begin{tikzpicture}[remember picture,overlay]
\node[anchor=south,yshift=10pt] at (current page.south) {\fbox{\parbox{\dimexpr\textwidth-\fboxsep-\fboxrule\relax}{\copyrighttext}}};
\end{tikzpicture}%
}    
    
\begin{document}

\title{ArCode: A Tool for Supporting Comprehension and Implementation of Architectural Concerns}

\maketitle
\copyrightnotice{}

\begin{abstract}
Integrated development environments (IDE) play an important role in supporting developers during program comprehension and completion. Many of these supportive features focus on low-level programming and debugging activities. Unfortunately, there is less support in understanding and implementing architectural concerns in the form of patterns, tactics and/or other concerns.
In this paper we present ArCode, a tool designed as a plugin for a popular IDE, IntelliJ IDEA. ArCode is able to learn correct ways of using frameworks' API to implement architectural concerns such as Authentication and Authorization. Analyzing the programmer's code, this tool is able to find deviations from correct implementation and provide fix recommendations alongside with graphical demonstrations to better communicate the recommendations with the developers. We showcase how programmers can benefit from ArCode by providing an API misuse detection and API recommendation scenario for a famous Java framework, Java Authentication and Authorization (JAAS) security framework.
\end{abstract}


\section{Introduction}
Nowadays program developers increasingly rely on Integrated Development Environments (IDEs) to accelerate the development pace, achieve better control over their projects, and implement programs that correctly follow the programming language syntax \cite{vihavainen2014novices, wu2020ggf}. These IDEs are supported and enhanced by enormous plugins and tools developed by experts to facilitate code implementations. 
IntelliJ IDEA\footnote{https://www.jetbrains.com/idea} is a leading IDE developed by JetBrains which comes with many features such as code completion, code navigation, and building projects \cite{intellij2011most}. Novice, as well as expert programmers receive many helpful fine-grained advice from these tools and plugins, mostly in a limited scope \cite{zayour2016qualitative, liu2020self}. For example, based on a partially written expression in a program, an auto-completion plugin would be able to leverage information obtained from the method context to provide some recommendations on how to complete the expression. These plugins make programmers’ life much easier, such that many program developers cannot write a simple program without them. 
While programmers benefit from plugins and tools provided by IDEs, there are some important concerns that have not been addressed by them. For instance, despite the advancements in programming languages and development environments, junior program developers still suffer from lack of proper comprehension of the program under development \cite{schroter2017comprehending, xia2017measuring, stapleton2020human, dias2020evaluating} including its architectural aspects \cite{galster2016makes}. Furthermore, they find it difficult to implement complicated usecases and architectural tactics from scratch \cite{harrison2007leveraging}. These tactics are mostly implemented by incorporating APIs of frameworks. Studies show that framework documentations are not well-prepared for junior programmers to learn how to implement different usecases by incorporating that framework in their code \cite{robillard2011field}. Moreover, some empirical studies show that programmers cannot rely on tutorials nor Q\&A posts to learn how to correctly incorporate these APIs for a tactic or usecase implementation \cite{baltes2019usage, fischer2017stack}. In these cases, even if a programmer receives fine-grained recommendations from the IDE, it would not be enough towards building a real-world functioning application.
Achieving this level of knowledge and experience and applying it in designing and implementing a program is not a trivial task. Therefore, to support programmers in the mentioned scenarios, empowering IDEs with more sophisticated tools and plugins is crucial. 

In this paper, we introduce ArCode, a tool that helps programmers understand implemented architectural concerns in their programs, find possible incorrect implementations, and fixing them. To that extent, Arcode learns how to correctly leverage API of frameworks from two sources: \textbf{(i)} by analyzing frameworks' bytecode, and \textbf{(ii)} from a sample of programs incorporating API of frameworks of interest. It then performs static analysis on the under development program and creates a graph-based model of API usage in that program. In an interaction with the programmer, ArCode is capable of providing recommendations on how to fix possible API misuses.

\section{ArCode}
\label{sec:ArCode}

ArCode is a tool for analyzing programs written in Java, creating their API usage models, identifying deviations in the code from a correct implementation, and providing recommendations on how to fix the found deviations. It is implemented as a plugin for IntelliJ IDEA development environment. There are three main inputs to ArCode:
\begin{itemize}
    \item Information about the framework of interest (e.g. JAAS framework)
    \item Path to training projects directory
    \item Path to the program under development
\end{itemize}

ArCode leverages static analysis techniques to extract some facts about API dependencies and their usage constraints from the framework binary code. It then analyzes programs in the training repository and filters out those that are violating API usage constraints. Mining the remained training programs, it builds a \texttt{Framework API Specification (FSpec)} model for the framework of interest. FSpec represents correct ways of leveraging a framework's API to implement usecases or architectural concerns. Next, ArCode analyzes the program under development and creates its graph-based API usage model. This model alongside with the created FSpec are used to compute the graph edit distance \cite{bunke1997relation} between the current implementation and a correct implementation. This distance shows how far the current implementation is from a correct way of implementing an architectural concern or usecase. If the computed distance is greater than zero, it means that there is a deviation from a correct implementation. Therefore, ArCode provides a ranked list of recommendations on how to fix the found deviation. Each recommendation consists of a graph-based demonstration of the correct version of API usage and an automatically generated code snippet that implements that correct API usage. 
Detailed scientific aspects of ArCode is presented in its technical paper \cite{shokri2021arcode}.

\subsection{Design and Implementation}
\begin{figure}
    \centering
        \includegraphics[width=.8\textwidth]{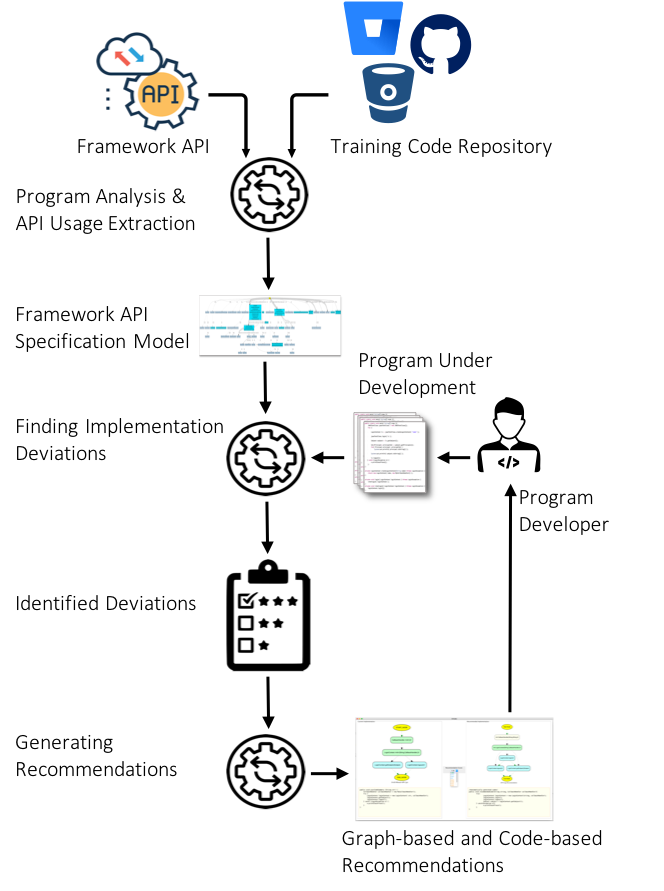}
        \centering
        \caption{ArCode main modules and their interactions.}
        \label{fig:ArCodeProcess}
\end{figure}

ArCode plugin is composed of three main modules which are responsible for \textbf{(i)} creating Framework API Specification (FSpec) model for a framework of interest, \textbf{(ii)} analyzing the program under development and identifying implementation deviations, and \textbf{(iii)} generating graph-based as well as code snippet recommendations. \figref{fig:ArCodeProcess} provides a high level overview of these modules and their interactions.  
\subsection{Features}
ArCode plugin helps programmers with three main features.
\subsubsection{Architectural Concern Summerization}
ArCode analyzes the program under development and creates API usage model for it. Then, a graph-based representation of this model is rendered and returned to the programmer. Moreover, a translation of this model to an executable java code snippet is automatically generated and provided for more comprehension of API usages in the program. 
\subsubsection{Implementation Deviation Detection}
After creating the API usage model of the program under development, ArCode searches through the created FSpec to identify incorrect API usages (i.e. API miuses) in the program. Then, it generates a list of change recommendations to be applied to the program. Each recommendation has a score value between 0 and 1 assigned to it. The higher the score is, the less changes needed to be applied to the program to make it a correct implementation. If score equals 1, it means that the program is following a correct implementation and the corresponding recommendation follows the same implementation. However, if there is no recommendation with score equal to 1, it means that the program deviates from a correct implementation and needs to be fixed. 
\subsubsection{Fix Recommendation}
In case that there is a deviation from correct implementation, programmer would be able to navigate between provided recommendations. Recommendations with higher scores are most likely to be what programmer intended to implement. Please not that a higher score indicates a higher similarity between the current implementation and a recommendation.  However, ArCode also provides recommendations with lower scores to give the programmer insights into what other correct API usages could be followed to make the code a correct implementation. Recommendation scores are colored green, blue, orange, and red to make programmers aware that the recommendation is very similar, somewhat similar, not similar, very different than the current implementation respectively.

\subsection{How to use ArCode}
\begin{figure}
    \centering
        \includegraphics[width=.8\textwidth]{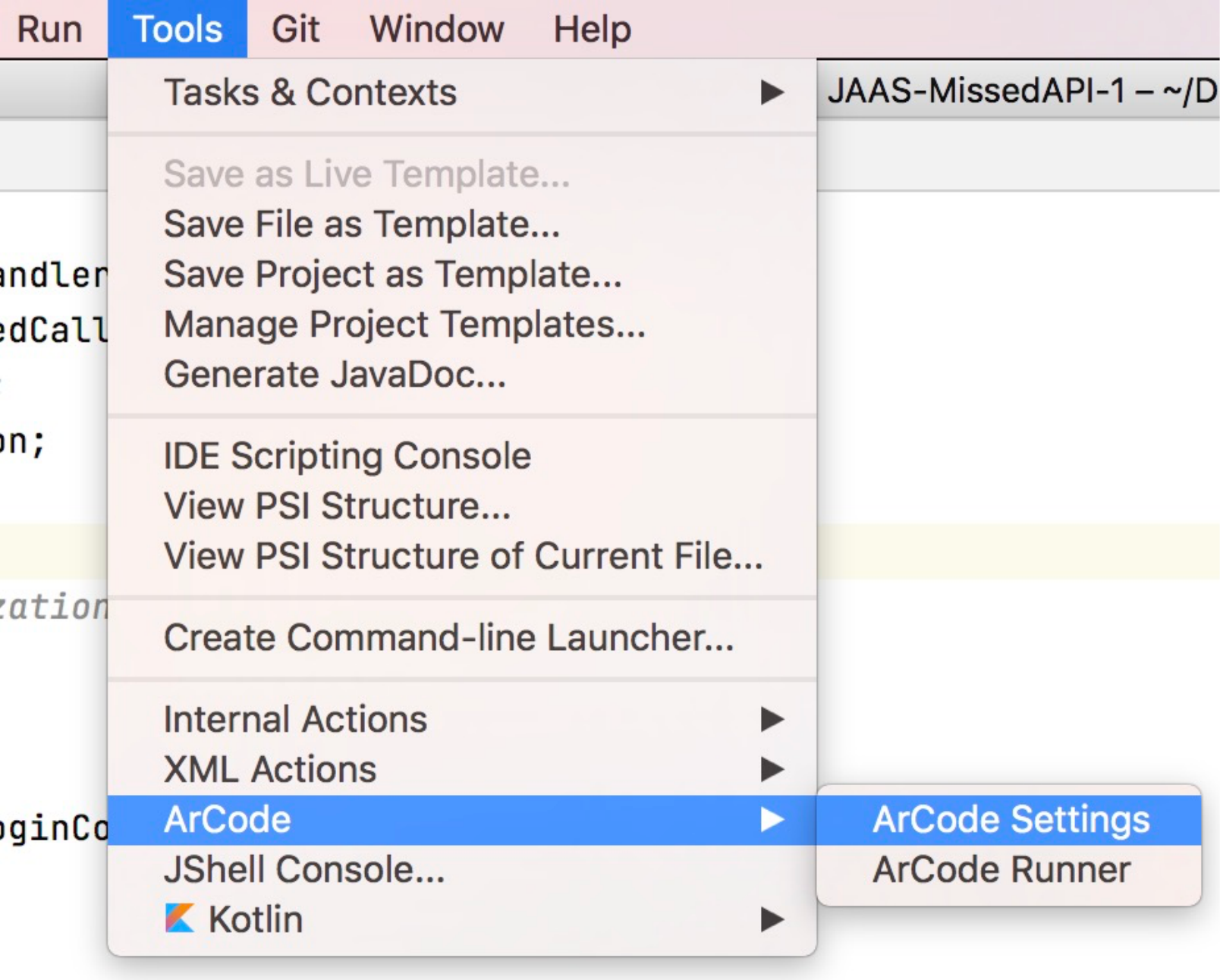}
        \centering
        \caption{ArCode main menu}
        \label{fig:ArCodeMenu}
\end{figure}

\begin{figure}
    \centering
        \includegraphics[width=.8\textwidth]{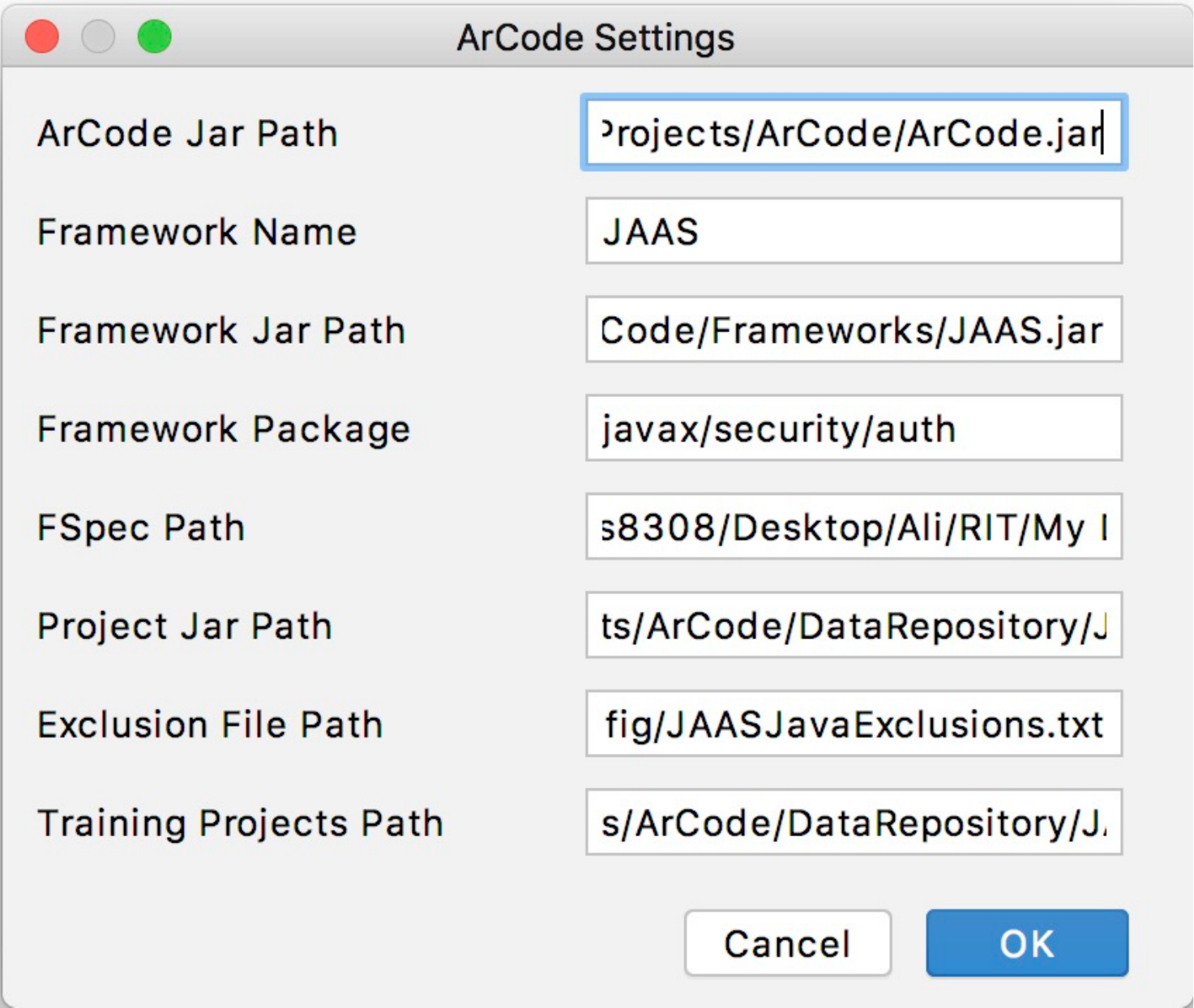}
        \centering
        \caption{ArCode Settings menu}
        \label{fig:ArCodeSettings}
\end{figure}

The source code as well as released versions of ArCode plugin are available at this repository on GitHub: \url{https://github.com/SoftwareDesignLab/ArCode-Plugin}. Once the plugin is installed on Intellij IDEA, ArCode's menu will appear as a sub-menu of \textit{tools} menu in this IDE (\figref{fig:ArCodeMenu}). Programmer needs to fill out the fields in the \textit{settings} menu as shown in \figref{fig:ArCodeSettings}. That is the way to configure ArCode plugin. More details on ArCode settings could be found from ArCode plugin repository. Then, by clicking on \textit{ArCode Runner} menu, ArCode starts analyzing the program and generating recommendations.    

\section{ArCode In Practice}
\label{sec:InPractice}
\begin{figure}
    \centering
        \includegraphics[width=.8\textwidth]{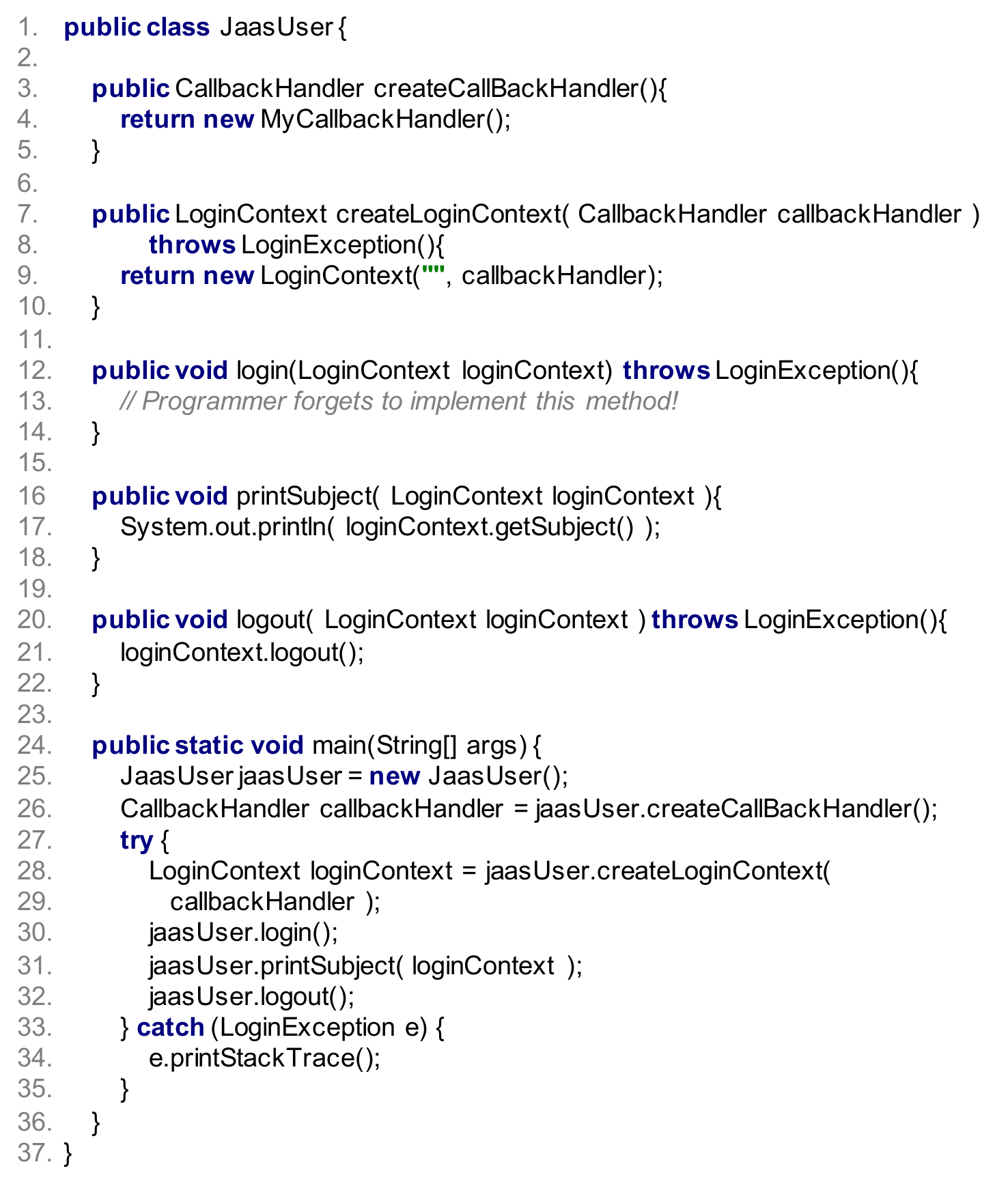}
        \centering
        \caption{A program that intends to implement authentication tactic.}
        \label{fig:CodeSample}
\end{figure}

To showcase how ArCode works, we start with a program in which a programmer aims to implement Authentication tactic by incorporating APIs from JAAS framework. \figref{fig:CodeSample} shows this program. The start point of the program is \code{main(...)} method. First, an object of \code{JaasUser} is created at line 25. This is a class that programmer has developed and is expected to correctly implement authentication tactic. Except for \code{main(...)} method, \code{JaasUser} has 5 methods each of which are responsible for calling an API of JAAS framework to support the authentication process. Following lines 26 to 32, programmer first instantiates an object of \code{LoginContext} at line 9 using an instantiated object of \code{CallbackHandler} from line 4. Then, it supposed to call \code{login()} method of instantiated \code{LoginContext} at line 13. However, the programmer mistakenly has not implemented this part. Next, the \code{Subject} that supposed to be populated by calling \code{login()} method is printed in line 17. Here, since \code{login()} method is not called, then \code{loginContext.getSubject()} at line 17 will return null and so it does not print the desired output (i.e. a semantical error). Finally, calling \code{logout()} method at line 21 concludes the authentication process. 

\begin{figure}
    \centering
    \begin{subfigure}[b]{0.8\textwidth}
        \centering
        \hspace{-1pt}
        \includegraphics[width=.8\textwidth]{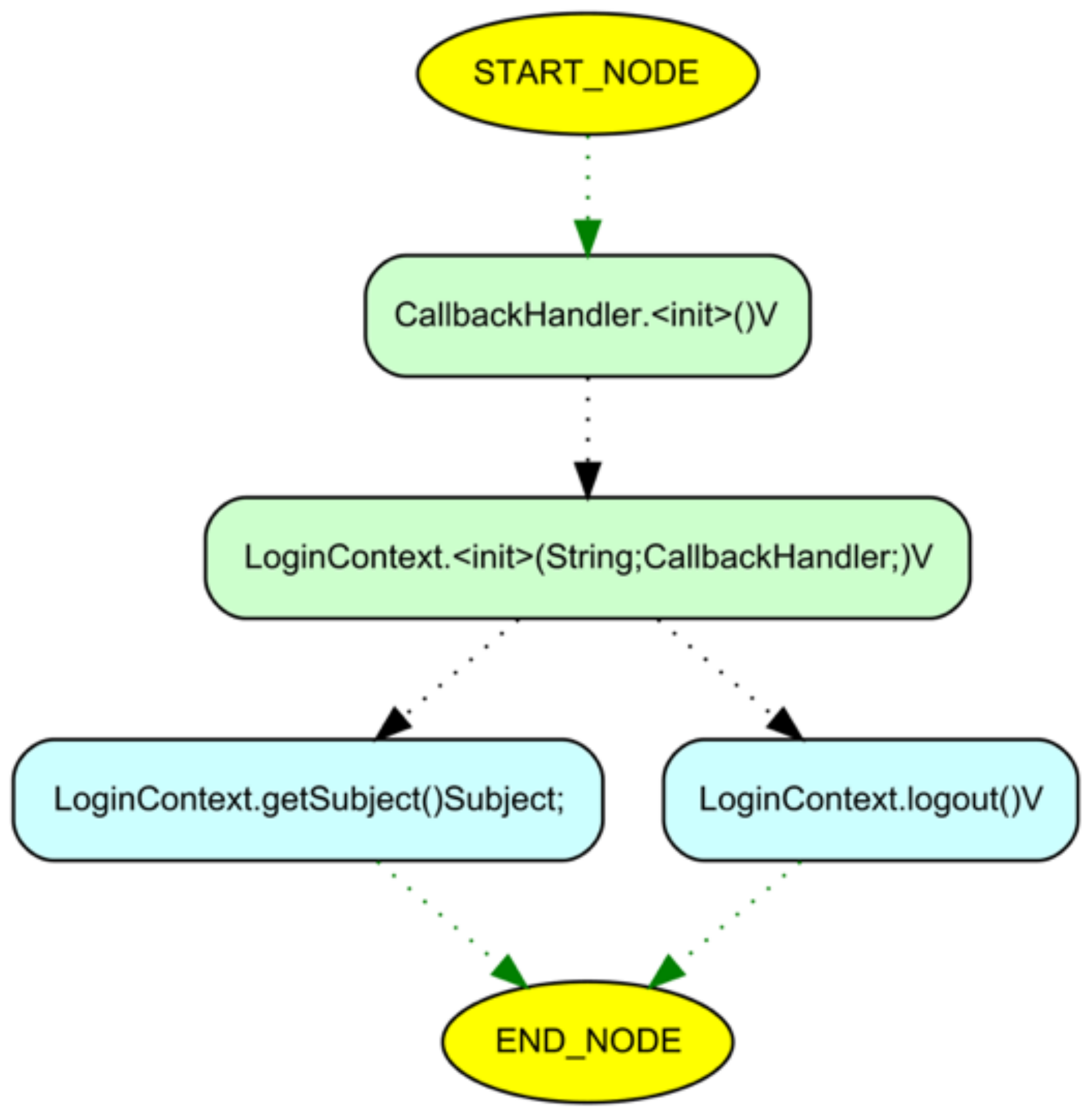}
        \centering
    \end{subfigure}%
 
    \begin{subfigure}[b]{.9\textwidth}
        \centering
        \includegraphics[width=\textwidth, right]{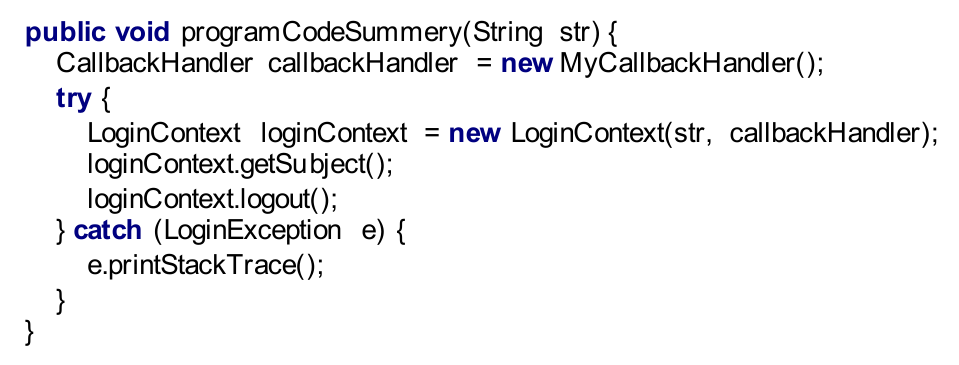}
    \end{subfigure}
        \caption{Summerized version of code snippet shown in \figref{fig:CodeSample} as a graph-based API usage model and a code snippet}
        \label{fig:CodeSampleSummery}
        \vspace{-6pt}

\end{figure}

This simple program demonstrates a possible case of API misuse when a programmer aims to implement architectural concerns. As APIs of JAAS framework are called in different methods, it would be hard for a programmer to correctly track API calls and their relationships. Specifically, tracking API usages in a program becomes a challenging and an error-prone task while working on a large-scale software. ArCode supports programmers with automatically tracing and summarizing these API usages and representing them as a graph-based API usage model. It also generates a summarized version of API usages as an executable Java code snippet. These summerizations help programmer better understand how architectural concerns are currently implemented in their program. \figref{fig:CodeSampleSummery} depicts the summerization (w.r.t. JAAS API) of the program shown in \figref{fig:CodeSample}.  

\begin{figure*}
    \centering
        \includegraphics[width=.8\textwidth]{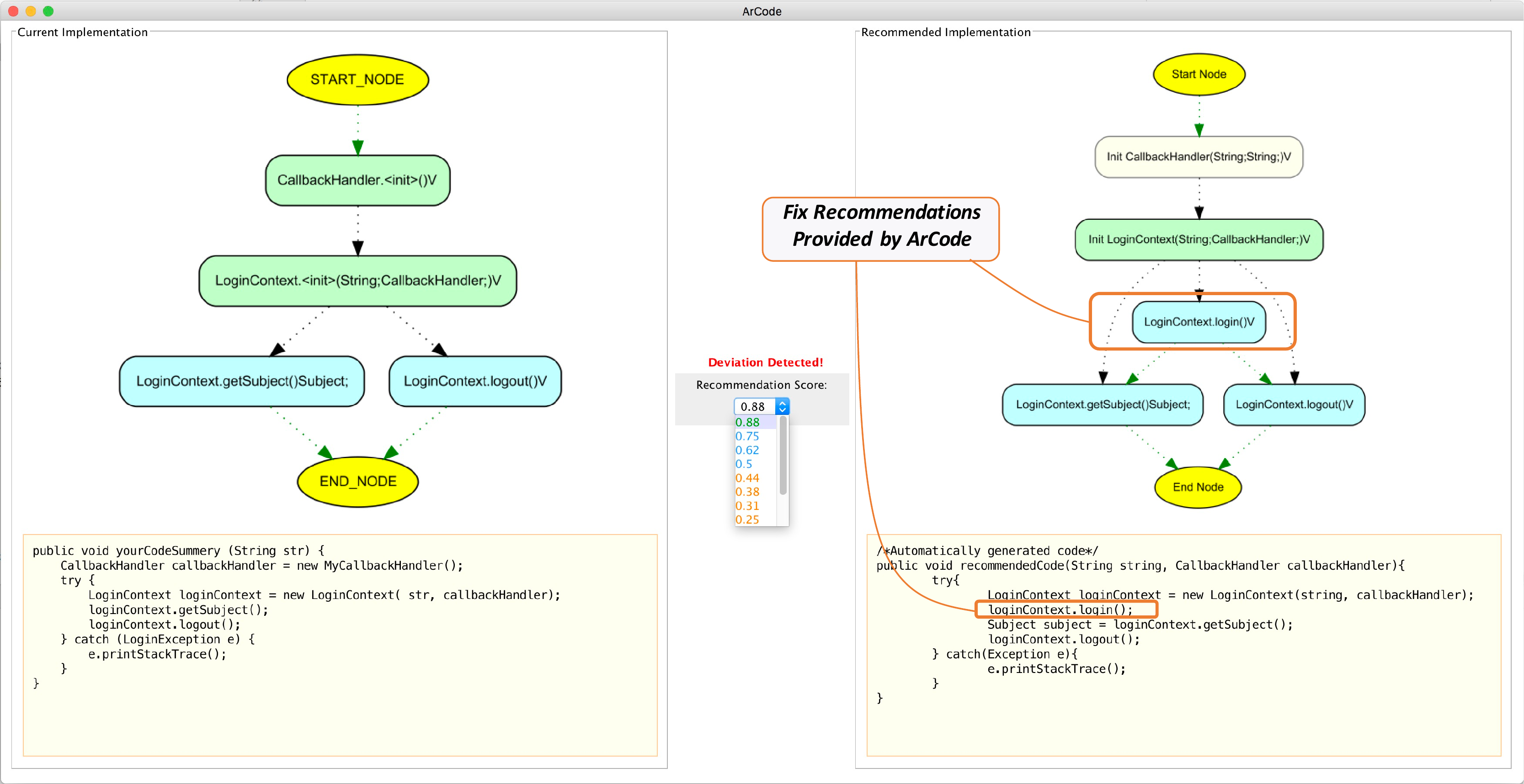}
        \centering
        \caption{Receiving recommendations from ArCode plugin in IntelliJ IDEA}
        \label{fig:ArCodeRec}
\end{figure*}

Next, ArCode leverages the built Framework API Specification (FSpec) model to identify deviations from a correct implementation. In case that such a deviation is found, a list of recommendations will be generated for the programmer.

\figref{fig:ArCodeRec} shows how ArCode plugin analyzes programmer's code, finds deviations from correct implementation, and provides recommendations. On the left side of this figure, a summarized version of the current implementation is shown as a graph-based API usage model as well as its related code snippet. On the right side, a possible way of fixing the found deviation is demonstrated by means of a graph-based API usage model and its corresponding code snippet. The programmer can navigate between the recommendation list through the items in the combo box. By selecting a new value from the combo box, the right panel will re-render the corresponding recommendation graph and code snippet. Values in the combo box represent the score of each recommendation. Scores are float numbers between 0 and 1. Recommendations with scores closer to 1 have more similarities to the current implementation. 
For programmers' convenient, scores in the range of [0.8, 1], [0.5, 0.8), [0.2, 0.5), and [0, 0.2) are colored as \textit{green}, \textit{blue}, \textit{orange}, and \textit{red} respectively. For instance, as shown in \figref{fig:ArCodeRec}, ArCode identifies that the programmer's current implementation deviates from a correct authentication implementation. Therefore, it provides a list of recommendations as a guidance on how to fix this problem. The highest score in the combo box of \figref{fig:ArCodeRec} is 0.88. It means that many parts of the program complies with a correct API usage. Nevertheless, there are some API misuses needed to be fixed. For instance, the recommended graph-based API usage model has an additional node, \code{LoginContext.login()}, compared to API usage model of the current implementation. Black dotted arrows represent data-dependencies between APIs. It means that programmer needs to call \code{login()} method of the instantiated \code{LoginContext} in the code. Green dotted arrows represent API call sequence constraints. It means that calling \code{login()} method must be placed in the code before the location that either \code{logout()} or \code{getSubject()} methods are called. Furthermore, on the lower right side of \figref{fig:ArCodeRec} an executable code snippet is displayed. Programmer can follow the code snippet to find a more comprehensive insight into the problem and its related fix recommendation. 

\section{Conclusion \& Future Work}
\label{sec:conclusion}
In this paper we introduced ArCode plugin, which is a tool that helps programmers achieve a better understanding of the current state of architectural concerns implementation in their program. It also facilitates finding and fixing implementation deviations. Although this tool provides very insightful fix recommendations, it is the programmer's responsibility to apply the suggested changes to correct the implementation. There might be a few interactions between the programmer and ArCode to make sure that the final code complies with a correct API usage.

As of our future work, we aim to automate the process of performing fix recommendations in the program under development. In that regard, based on the selected recommendation, ArCode will automatically change the program. Then, it will re-analyze the code, re-generate program's API usage model and code summery, and populates a new list of recommendations. Also, future work direction includes automatically synthesizing the code from scratch based on the selected API usage by the programmer. 

\section*{Acknowledgments}
This work was partially funded by the US National Science Foundation (NSF) under grant number CCF-1943300, CNS-1816845 and CNS-1823246. 

\bibliographystyle{IEEEtran}
\bibliography{IEEEabrv,bibliography}

\end{document}